\documentclass[letterpaper,english,reprint,nofootinbib,aps,superscriptaddress,showpacs,showkeys,prl]{revtex4-1}

\usepackage{babel,calc,amsmath,amsthm,amssymb,graphicx,subfigure,xcolor,comment}
\usepackage{txfonts}
\usepackage{mathdots}
\usepackage{todonotes}
\usepackage[T1]{fontenc}
\usepackage[unicode=true]{hyperref}
\usepackage{booktabs}
\usepackage{threeparttable}
\usepackage{times}
\usepackage{CJK}
\hypersetup{
	colorlinks=true,       		% false: boxed links; true: colored links
	linkcolor=blue,          	% color of internal links
	citecolor=blue,            % color of links to bibliography
	urlcolor=black,           	% color of external links
}

\theoremstyle{definition}

\def\ket#1{\left\lvert {#1} \right\rangle}

\begin{document}

\begin{CJK*}{UTF8}{gbsn}

\title{Hong-Ou-Mandel Interference between Two Hyper-Entangled Photons Enables Observation of Symmetric and Anti-Symmetric Particle Exchange Phases}

\author{Zhi-Feng Liu (刘志峰)}
\author{Chao Chen (陈超)}
\affiliation{National Laboratory of Solid State Microstructures, School of Physics, Nanjing University, Nanjing 210093, China}
\affiliation{Collaborative Innovation Center of Advanced Microstructures, Nanjing 210093, China}

\author{Jia-Min Xu (徐佳敏)}
\affiliation{Hefei National Laboratory, Hefei 230088, China}
\affiliation{Synergetic Innovation Center of Quantum Information and Quantum Physics, University of Science and Technology of China, Hefei 230026, China}

\author{Zi-Mo Cheng (程子默)}
\author{\\Zhi-Cheng Ren (任志成)}
\author{Bo-Wen Dong (董博文)}
\author{Yan-Chao Lou (娄严超)}
\author{\\Yu-Xiang Yang (杨雨翔)}
\author{Shu-Tian Xue (薛舒天)}
\author{Zhi-Hong Liu (刘志红)}
\author{\\Wen-Zheng Zhu (朱文正)}
\affiliation{National Laboratory of Solid State Microstructures, School of Physics, Nanjing University, Nanjing 210093, China}
\affiliation{Collaborative Innovation Center of Advanced Microstructures, Nanjing 210093, China}

\author{Xi-Lin Wang (汪喜林)}
\email[]{xilinwang@nju.edu.cn}
\affiliation{National Laboratory of Solid State Microstructures, School of Physics, Nanjing University, Nanjing 210093, China}
\affiliation{Collaborative Innovation Center of Advanced Microstructures, Nanjing 210093, China}
\affiliation{Hefei National Laboratory, Hefei 230088, China}
\affiliation{Synergetic Innovation Center of Quantum Information and Quantum Physics, University of Science and Technology of China, Hefei 230026, China}

\author{Hui-Tian Wang (王慧田)}
\email[]{htwang@nju.edu.cn}
\affiliation{National Laboratory of Solid State Microstructures, School of Physics, Nanjing University, Nanjing 210093, China}
\affiliation{Collaborative Innovation Center of Advanced Microstructures, Nanjing 210093, China}
\affiliation{Collaborative Innovation Center of Extreme Optics, Shanxi University, Taiyuan 030006, China}

%\date{\today}

%%%%%%%%%%%%%%%%%%%%%%%%%%%%%%%%%%%

\begin{abstract}
Two-photon Hong-Ou-Mandel (HOM) interference is a fundamental quantum effect with no classical counterpart. The exiting researches on two-photon interference were mainly limited in one degree of freedom (DoF), hence it is still a challenge to realize the quantum interference in multiple DoFs. Here we demonstrate the HOM interference between two hyper-entangled photons in two DoFs of polarization and orbital angular momentum (OAM) for all the sixteen hyper-entangled Bell states. We observe hyper-entangled two-photon interference with bunching effect for ten symmetric states (nine Boson-Boson states, one Fermion-Fermion state) and anti-bunching effect for six anti-symmetric states (three Boson-Fermion states, three Fermion-Boson states). More interestingly, expanding the Hilbert space by introducing an extra DoF for two photons enables to transfer the unmeasurable external phase in the initial DoF to a measurable internal phase in the expanded two DoFs. We directly measured the symmetric exchange phases being $0.012 \pm 0.002$, $0.025 \pm 0.002$ and $0.027 \pm 0.002$ in radian for the three Boson states in OAM and the anti-symmetric exchange phase being $0.991 \pi \pm 0.002$ in radian for the other Fermion state, as theoretical predictions. Our work may not only pave the way for more wide applications of quantum interference, but also develop new technologies by expanding Hilbert space in more DoFs. 
\end{abstract}
	
\maketitle

\end{CJK*}

%%%%%%%%%%%%%%%%%%%%%%%%%%%%%%%%%%%

Two-photon interference, Hong-Ou-Mandel (HOM) interference~\cite{Hong1987}, is a genuine quantum effect with no classical counterpart. Since a single experiment can unveil the quantum features of particle-wave duality and indistinguishability simultaneously, the HOM interference has attracted significant interest over the past decades~\cite{Mandel1999, Bouchard2020} and has been generalized to many other quantum particles or quasi-particles, such as two atoms~\cite{Kaufman2014, Lopes2015}, two deterministic collective excitations in an atomic ensemble~\cite{Li2016}, two surface plasmons~\cite{Heeres2013,Fakonas2014, Martino2014, Cai2014} and two phonons~\cite{Toyoda2015}. Unlike the bunching effect in the HOM interference of two indistinguishable Bosons, the anti-bunching effect will be resulted from the interference of two indistinguishable Fermions~\cite{Liu1998,Bocquillon2013}.

The HOM interference has become a cornerstone of modern quantum technologies and widely utilized to characterize the single photons from solid-state emitter~\cite{He2013, Ding2016, Somaschi2016, Wang2019, Tomm2021} and the photon pairs from spontaneous parametric down-conversion (SPDC)~\cite{Wang2016, Zhong2018, Zhong2020}. The HOM interference is also an important way to combine photons to implement controlled-NOT gate~\cite{O'Brien2003, Bao2007, Li2021} and to construct multi-photon entangled states including N00N states~\cite{Walther2004, Nagata2007}, GHZ states~\cite{Pan2012, Wang2016, Zhong2018}, graph states~\cite{Lu2007, Pan2012} and multi-photon high-dimensional entangled states~\cite{Malik2016, Zhang2017, Erhard2018}. The HOM interference is also the foundation of implementing the Bell state measurement, which is crucial for a variety of important quantum protocols, such as quantum teleportation~\cite{Bouwmeester1997, Wang2015, Pirandola2015, Ren2017}, entanglement swapping~\cite{Pan1998, Basset2019, Zopf2019} and connecting quantum nodes in quantum network~\cite{Hofmann2012, Hensen2015, Yu2020, Liu2021}.

Two-photon interference in more general scenarios has attracted significant interest, such as high dimensions~\cite{Zhang2016, Hiekkamaki2021}, multiple-mode states~\cite{Walborn2003, Francesconi2020}, and structured light~\cite{D'Ambrosio2019}. The HOM interference can be utilized as a state filter, which can engineer two-photon high-dimensional states~\cite{Zhang2016}. The HOM interference is determined by the exchange symmetry of two-photon states, where the symmetric (anti-symmetric) state leads to the bunching (anti-bunching) effect. So the key is how to harness the symmetry of two-photon state for the HOM interference. One well-known method is to prepare the symmetric and anti-symmetric two-photon states with the Bell states, as all the Bell states have the symmetry and form a set of orthogonal and complete bases. Another interesting method is to tame the symmetry of two-photon state by tailoring the multiple mode. For example, the multi-mode HOM interference was realized by transferring the symmetry of transverse spatial modes from the pump to the produced two-photon state via the SPDC and the global symmetry of two-photon state was further controlled by the polarization DoF (only one of the four polarization Bell states is anti-symmetric)~\cite{Walborn2003}. Recently, one has reported that the symmetry of two-photon state in frequency DoF can be controlled via the spatial mode of the pump~\cite{Francesconi2020}. It has been revealed that the hidden entanglement is in the anti-symmetric state~\cite{Fedrizzi2009}, which gives rise to the observation of anti-bunching HOM effect. 

Using the hyper-entangled photons in multiple DoFs in the Bell-state bases, the HOM interference can be explored in a more general case, e.g., changing the symmetry of two-photon state by selecting the different Bell states in each DoF independently. However, most previous experiments were limited to the Bell-state entanglement in one DoF. It is still a challenge to experimentally explore the HOM interference in two and more DoFs. Here we demonstrate the two-photon HOM interference in two DoFs of polarization and orbital angular momentum (OAM) in all the sixteen hyper-entangled Bell states. Moreover, based on the HOM interference in two DoFs, we successfully measure the exchange phases of two photons in the OAM DoF for the four OAM Bell states including three symmetric and one anti-symmetric exchange phases. The crux to success is to extend the Hilbert space by introducing an extra DoF of polarization to convert the unmeasurable external exchange phase in one OAM DoF into the measurable internal phase in two DoFs of polarization and OAM. 

\begin{figure}
\centering
\includegraphics[width=\linewidth]{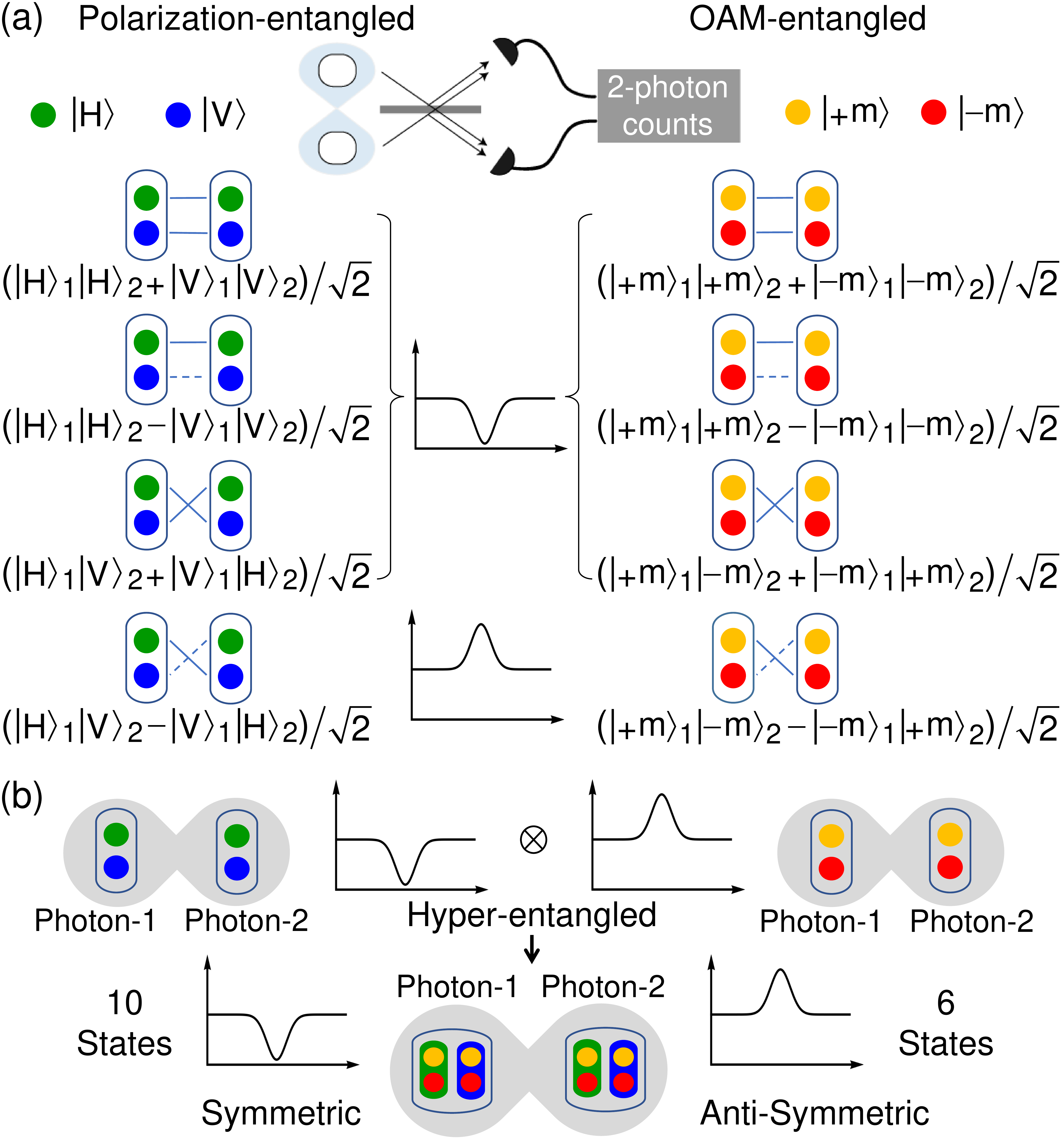}
 \caption{Two-photon HOM interference. (a) Four Bell states in one DoF of polarization or OAM containing three exchange symmetric Boson states and one exchange anti-symmetric Fermion state. (b) Polarization-OAM hyper-entangled states containing10 symmetric and 6 anti-symmetric states.}
\label{fig:1}
\end{figure}

In one DoF of polarization, the four Bell states perfectly characterize the exchange symmetries of two photons~\cite{Sun2007}, $\ket{\phi^{\pm}}_{12}$ and $\ket{\psi^{\pm}}_{12}$, can be written as
\begin{align} 
&\ket{\phi^{\pm}}_{12}=(\ket{H}_{1}\ket{H}_{2} \pm \ket{V}_{1}\ket{V}_{2})/\sqrt{2}, \\
&\ket{\psi^{\pm}}_{12}=(\ket{H}_{1}\ket{V}_{2} \pm \ket{V}_{1}\ket{H}_{2})/\sqrt{2},
\end{align}
where $\ket{H}$ ($\ket{V}$) refers to horizontal (vertical) linearly polarized state. Three Bell states $\ket{\phi^{\pm}}_{12}$ and $\ket{\psi^{+}}_{12}$ (one $\ket{\psi^{-}}_{12}$) have exchange symmetry (anti-symmetry) with the feature of $\ket{\varphi}_{21}=\ket{\varphi}_{12}$ ($\ket{\varphi}_{21}=-\ket{\varphi}_{12}$) and exhibit the bunching (anti-bunching) effect, i.e. a central dip (peak) in the HOM interference curve shown in Fig.~\ref{fig:1}(a), so the three states (the other state) belong to Boson states (Fermion state). In fact, any identical two photons with the state of $\ket{\varphi}_{12}=(\alpha \ket{H}_{1} + \beta \ket{V}_{1}) \otimes (\alpha \ket{H}_{2} + \beta \ket{V}_{2})$ (where $\alpha$ and $\beta$ are normalized complex constants satisfying $\left| \alpha \right|^2+\left| \beta \right|^2=1$) can be represented as the superposition of the three Boson states, resulting in that two indistinguishable photons are exchange symmetric, as the HOM interference between two identical photons. Therefore, the four Bell states are essential to describe the bunching and anti-bunching effects in the two-photon interference.    

Similarly, we can remark the four Bell states in the DoF of OAM, $\ket{\mu^{\pm}}_{12}$ and $\ket{\nu^{\pm}}_{12}$ as follows
\begin{align}
\ket{\mu^{\pm}}_{12} & = (\ket{+m}_{1}\ket{+m}_{2} \pm \ket{-m}_{1}\ket{-m}_{2})/\sqrt{2}, \\
\ket{\nu^{\pm}}_{12} & = (\ket{+m}_{1}\ket{-m}_{2} \pm \ket{-m}_{1}\ket{+m}_{2})/\sqrt{2},
\end{align}
where $\ket{\pm m}$ refer to the photon states with OAMs of $ \pm m\hbar$ per photon, respectively. The four OAM Bell states can also be divided into two categories of Boson states ($\ket{\mu^{\pm}}_{12}$ and $\ket{\nu^{+}}_{12}$) and Fermion state ($\ket{\nu^{-}}_{12}$), leading to three central dips and one central peak in the HOM interference curves shown in Fig.~\ref{fig:1}(a). When exploring two-photon interference in two DoFs of polarization and OAM, a hyper-entangled state (Boson/Fermion-Boson/Fermion state) in the form of the product of a polarization Bell state and an OAM Bell state can be prepared for two photons, as shown in Fig.~\ref{fig:1}(b). Interesting conclusions would be expected for the two-photon interference with hyper-entangled state according to the general physics law that more is different~\cite{Anderson1972}.

To explore the two-photon HOM interference in two DoFs, we classify sixteen hyper-entangled states into four groups, i.e., one Fermion-Fermion state ($\ket{\psi^{-}}\otimes\ket{\nu^{-}}$), three Fermion-Boson states ($\ket{\psi^{-}}\otimes\{\ket{\mu^{+}},\ket{\mu^{-}},\ket{\nu^{+}}\}$), three Boson-Fermion states ($\{\ket{\phi^{+}},\ket{\phi^{-}},\ket{\psi^{+}}\}\otimes\ket{\nu^{-}}$), and  nine Boson-Boson states ($\{\ket{\phi^{+}},\ket{\phi^{-}},\ket{\psi^{+}}\}\otimes\{\ket{\mu^{+}},\ket{\mu^{-}},\ket{\nu^{+}}\}$). The exchange symmetry governing the bunching or anti-bunching effect in the HOM interference can be revealed by the number of Fermion states appearing in the hyper-entangled state. When a hyper-entangled state contains odd Fermion states, it is exchange anti-symmetric and results in the anti-bunching effect in the HOM interference. As a comparison, even Fermion states appearing in a hyper-entangled state will lead to the exchange symmetry and the bunching effect. This conclusion is scalable to hyper-entangled state in the more DoFs.

\begin{figure*}[!htb]
	\centering
	\includegraphics[width=0.9\linewidth]{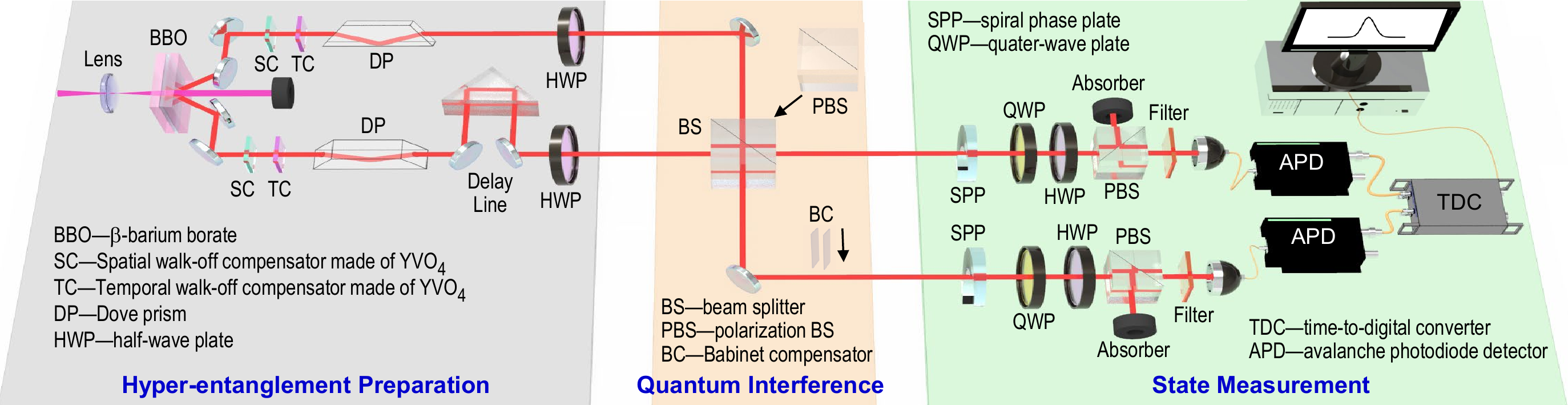}
	\caption{Experimental setup for hyper-entangled two-photon interference on a BS and direct measurement of two-photon exchange phase, where a PBS replaces the interference element to realize two-photon exchange. We have added mirror(s) in some optical paths to ensure that any OAM state is reflected for even times, so as to remain its handedness unchanged.}
	\label{fig:2}
\end{figure*} 

\begin{figure*}[!tb]
	\centering
	\includegraphics[width=0.8\linewidth]{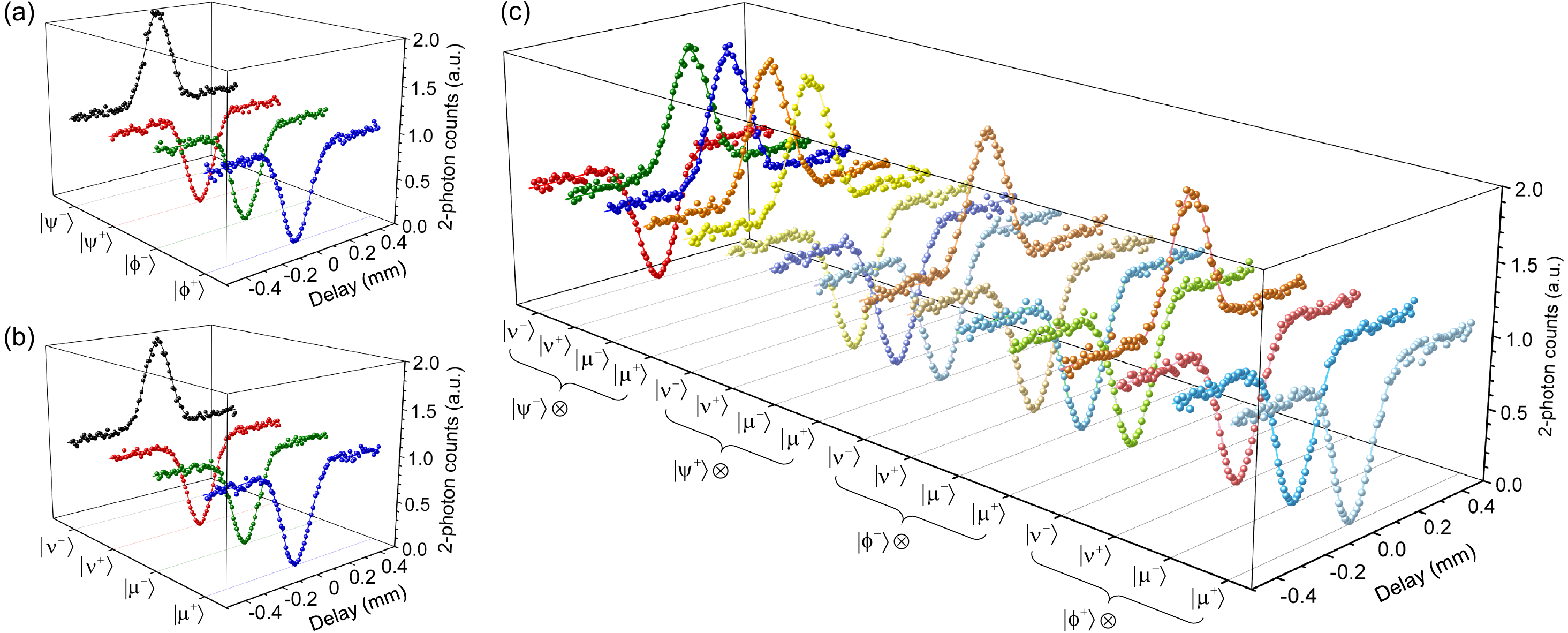}
	\caption{HOM interference when using a pair of filters with 3-nm bandwidth for (a) polarization Bell states, (b) orbital angular momentum Bell states, and (c) hyper-entangled Bell states. The error bars are concealed as they are smaller than the data spheres.}
	\label{fig:3}
\end{figure*}
In the experimental setup (Fig.~\ref{fig:2}), the hyper-entangled photon pairs are generated via the SPDC when a femtosecond (fs) pulsed laser passes through two 0.6-mm-thick type-I BBO ($\beta$-barium borate) crystals glued with orthogonally configured optic axes~\cite{Barreiro2005, Wang2015}, in which the polarization and OAM entanglements are achieved simultaneously in the small-angle type-I SPDC~\cite{Mair2001}. The spatial and temporal walk-offs between photons from two BBO crystals are compensated with various YVO$_{4}$ crystals (see~\cite{SuppMat} for details). Thus we directly prepare the hyper-entangled state $\ket{\phi^{+}}\otimes\ket{\nu^{+}}$ and the other fifteen hyper-entangled states can also be easily realized with the aid of two Dove prisms (DPs) and two half wave plates (HWPs) (see~\cite{SuppMat} for details). Here the prepared OAM entanglement in the subspace has the OAMs of $\pm \hbar$ (i.e. $m = 1$). Then the two photons enter into two input ports of a beam splitter (BS) to implement the HOM interference in two DoFs, where the arriving time of one photon is controlled by the delay line. A state measurement configuration by recording the two-photon coincidence counts in the two output port of BS is utilized to obtain the HOM interference curve between two hyper-entangled photons. For the first one $\ket{\phi^{+}}\otimes\ket{\nu^{+}}$ among the hyper-entangled states, we choose $\ket{H}_{1} \ket{+1}_{1}$ and $\ket{H}_{2} \ket{-1}_{2}$ as the two-photon measurement bases (see~\cite{SuppMat} for all the sixteen measurement bases). 

When a pair of narrow-band filters with 3-nm bandwidth is utilized, the measured HOM interference in one DoF of polarization or OAM in Fig.~\ref{fig:3}(a) or (b) shows three dips at zero delay for the three Boson states and one peak at zero delay for the Fermion state. As shown in Fig.~\ref{fig:3}(c), the HOM interference in the polarization-OAM hyper-entangled states exhibits ten dips in nine Boson-Boson states ($\{\ket{\phi^{+}},\ket{\phi^{-}},\ket{\psi^{+}}\}\otimes\{\ket{\mu^{+}},\ket{\mu^{-}},\ket{\nu^{+}}\}$) and one Fermion-Fermion state ($\ket{\psi^{-}}\otimes\ket{\nu^{-}}$), and six peaks in three Boson-Fermion states ($\{\ket{\phi^{+}},\ket{\phi^{-}},\ket{\psi^{+}}\}\otimes\ket{\nu^{-}}$) and three Fermion-Boson states ($\ket{\psi^{-}}\otimes\{\ket{\mu^{+}},\ket{\mu^{-}},\ket{\nu^{+}}\}$, as the theoretical predictions. To characterize a HOM interference, we utilize the visibility defined as $V_{dip} = 1  - C_{0}/C_{\infty}$ ($V_{peak} = C_{0}/C_{\infty} - 1$) for the dip (peak)~\cite{Wang2015,Gao2022}, where $C_{0}$ and $C_{\infty}$ are fitted counts at zero and infinite delays. The extracted visibilities of the HOM interference for the four Bell states in the single DoF of polarization or OAM ranges from $0.975 \pm 0.005$ to $0.992 \pm 0.066$ or $0.957 \pm 0.068$ to $0.993 \pm 0.003$; while the HOM interference for the sixteen hyper-entangled states has the extracted visibilities ranging from $0.902 \pm 0.071$ to $0.993 \pm 0.002$ (see~\cite{SuppMat} for details and the corresponding results with 8-nm filters).

An interesting application for the hyper-entangled two-photon interference is the direct measurement of the two-photon exchange phases. Recently, an approach has been in theory proposed to reveal the quantum statistics by constructing the superposition of a reference state of two distant particles and a physically exchanged two-particle state~\cite{Roos2017}. The experiment has also demonstrated the direct measurement of exchange phase of indistinguishable photons, which is the symmetric exchange phase for two-photon Boson state~\cite{Tschernig2021}. In fact, expanding the Hilbert space by introducing an extra DoF for the two photons enables to transfer the unmeasurable external (global) phase in a single DoF to a measurable internal phase in the expanded two DoFs. 

\begin{figure}[!htb]
	\centering
	\includegraphics[width=0.95\linewidth]{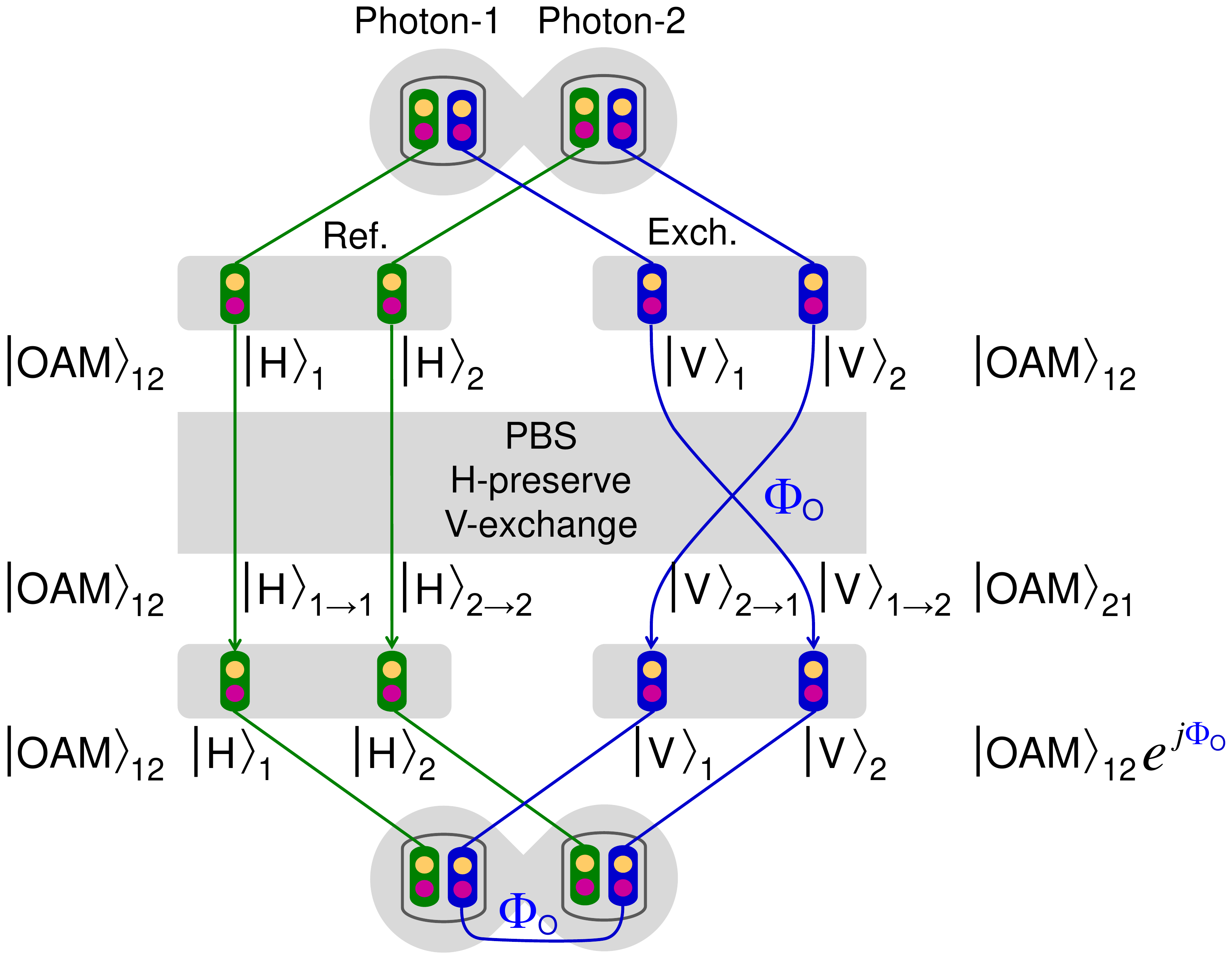}
	\caption{Principle to directly measure the exchange phases for two photons in the OAM state by the aid of polarization as an ancillary DoF with hyper-entangled two-photon interference. The exchange process is realized on PBS for vertically polarized component.}
	\label{fig:4}
\end{figure}

Figure~\ref{fig:4} shows the principle for directly measuring the exchange phases for the two-photon OAM entangled states. The exchange process can be
described as $\ket{OAM}_{12} \xrightarrow{\rm Exchange} \ket{OAM}_{21} = e^{j \Phi_O} \ket{OAM}_{12}$. In theory, the OAM exchange phase $\Phi_O$ is zero for a Boson state and $\pi$ for the Fermion state. In the four OAM Bell states $\ket{\mu^{\pm}}_{12}$ and $\ket{\nu^{\pm}}_{12}$, the three Boson states of $\ket{\mu^{\pm}}_{12}$ and $\ket{\nu^{+}}_{12}$ will result in a zero exchange phase. It is worth noting that, two identical photons is in the superposition state of these three Boson states, therefore it is also a Boson state and has a zero exchange symmetric phase. For two photons, only the unique Fermion OAM state of $\ket{\nu^{-}}_{12}$ can result in an exchange anti-symmetric phase of $\pi$.

To directly measure these exchange phases, we introduce an extra DoF of polarization and then prepare the initial hyper-entangled state of $\ket{\phi^{+}}_{12} \otimes \ket{OAM}_{12}$ in the two DoFs of polarization and OAM as
\begin{equation}
 (\ket{H}_{1}\ket{H}_{2} + \ket{V}_{1} \ket{V}_{2})/\sqrt{2} \otimes \ket{OAM}_{12},
\end{equation} 
where $\ket{H}_{1}\ket{H}_{2} \otimes \ket{OAM}_{12}$ is considered as the reference two-photon state and the two-photon state of $\ket{V}_{1} \ket{V}_{2} \otimes \ket{OAM}_{12}$ will be exchanged. When such a two-photon state passes through a PBS, the transmitted horizontal polarization component will be physically preserved on its original propagation path as the input one, therefore, this polarization component could be utilized as the the reference state. On the contrary, for the vertical polarization component in the two-photon state of $\ket{V}_{1}\ket{V}_{2}\otimes\ket{OAM}_{12}$, it will be reflected on the PBS and the two photons will be physically exchanged, resulting in the exchange phase $\Phi_O$ with respect to its original two-photon state, which will follow the process below
\begin{equation}
\begin{aligned}
\ket{V}_{1}\ket{V}_{2}\otimes \ket{OAM}_{12} & \xrightarrow[\rm by \  PBS]{\rm Exchange} e^{j \pi}\ket{V}_{2}\ket{V}_{1}\otimes \ket{OAM}_{21}\\
&=e^{j (\pi+\Phi_P)}\ket{V}_{1}\ket{V}_{2}\otimes e^{j \Phi_O}\ket{OAM}_{12} \\
& \xrightarrow{\rm BC} \ket{V}_{1}\ket{V}_{2}\otimes e^{j \Phi_O}\ket{OAM}_{12}.
\end{aligned}
\end{equation}
Considering only the DoF of polarization, for the two-photon exchange process on the PBS, the reflections will introduce a $\pi$ phase for $\ket{V}_{2}\ket{V}_{1}$. In addition, an exchange phase $ \Phi_P$ for the polarization will be also brought as $\ket{V}_{2}\ket{V}_{1}=e^{j \Phi_P} \ket{V}_{1}\ket{V}_{2}$. The total phase of $(\pi+\Phi_P)$ appears in the DoF of polarization, and can be compensated by a Babinet compensator (BC) before measuring the OAM exchange phase $\Phi_O$.

After the PBS and the BC, the OAM exchange phase will appear in the two-photon vertical polarization component, yielding the superposition state 
\begin{equation}
(\ket{H}_{1}\ket{H}_{2} + e^{j \Phi_O} \ket{V}_{1}\ket{V}_{2})/\sqrt{2}\otimes\ket{OAM}_{12}.
\end{equation}
The measurement of external phase $\Phi_O$ for the exchanged two-photon OAM state $e^{j \Phi_O} \ket{OAM}_{12}$ has been turned into the measurement of internal phase in the two-photon polarization state $(\ket{H}_{1}\ket{H}_{2} + e^{j \Phi_O} \ket{V}_{1}\ket{V}_{2})/\sqrt{2}$ between $\ket{H}_{1}\ket{H}_{2}$ and $ \ket{V}_{1}\ket{V}_{2}$. In experiment, the above process can be easily implemented by replacing the BS with a PBS and inserting a BC into the quantum interference unit shown in Fig.~\ref{fig:2}. 

To measure $\Phi_O$ in experiment, for the two-photon polarization state $(\ket{H}_{1}\ket{H}_{2} + e^{j \Phi_O} \ket{V}_{1}\ket{V}_{2})/\sqrt{2}$, we require to project the photon-1 into the basis of $(\ket{H}_{1} + \ket{V}_{1})/\sqrt{2}$ and then to measure the photon-2 in the basis of $(\ket{H}_{2} \pm e^{j\theta}\ket{V}_{2})/\sqrt{2}$ that are the eigenstates of the observable $M_{\theta}=\cos \theta \sigma_{x} + \sin \theta \sigma_{y}$ with their eigenvalues of $+1$ and $-1$, where $\sigma_{x}$ and $\sigma_{y}$ are the Pauli $x$ and $y$ matrices. Here $\theta$ spans from 0 to $2\pi$, in experiment $\theta$ is determined by the orientation angle $\Theta$ of the HWP's fast axis with respect to the horizontal direction, where $\Theta = 3\pi/8+\theta/4$ (see~\cite{SuppMat} for details).

\begin{figure}[!tb]
	\centering
	\includegraphics[width=0.8\linewidth]{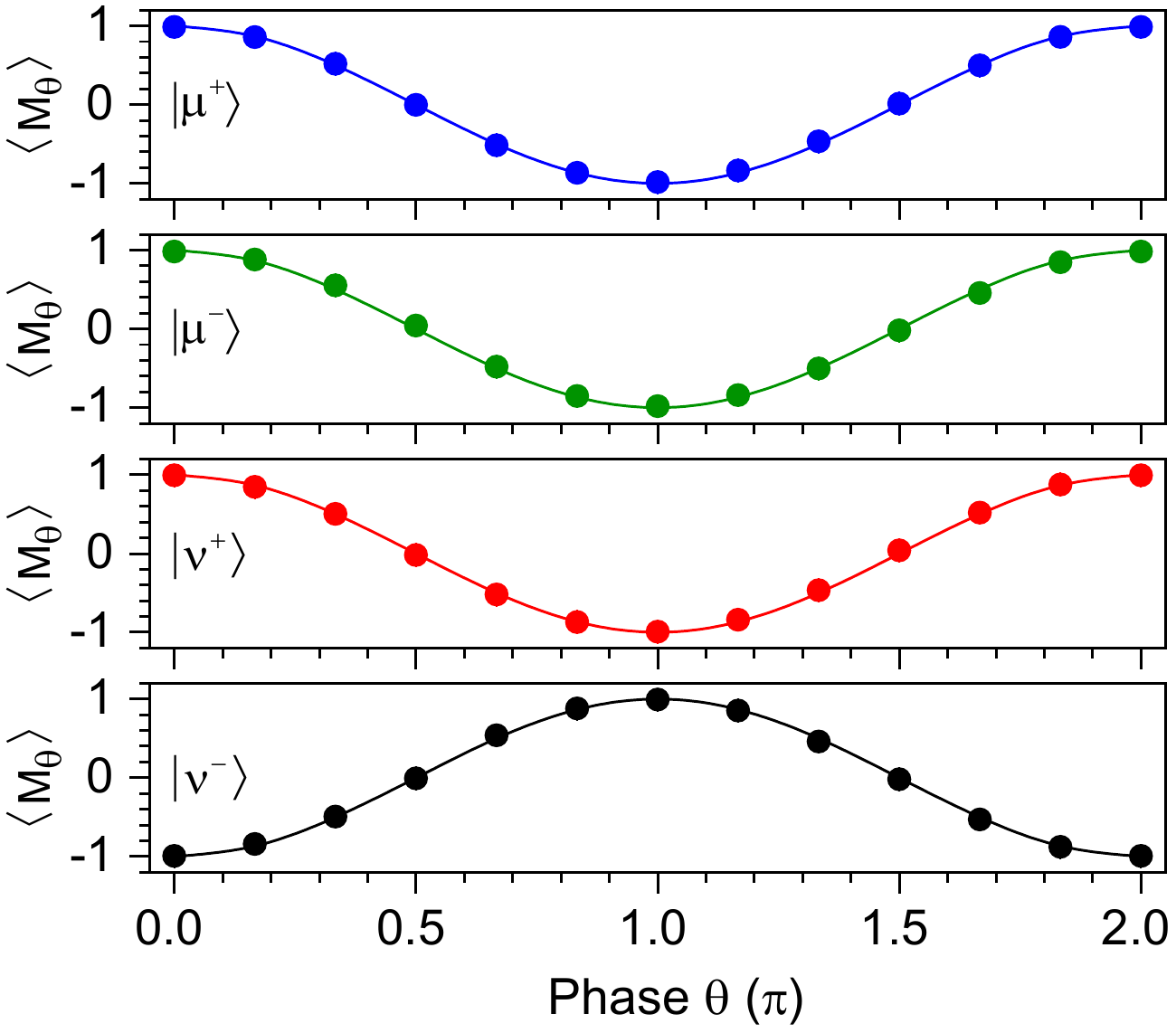}
	\caption{Experimental results for exchange phase of two photons in the four OAM Bell states. Two-photon counts are recorded by projecting the photon-1 in the basis of $(\ket{H}+\ket{V})/\sqrt{2}$ and measuring the photon-2 in the basis of $(\ket{H} \pm e^{j\theta}\ket{V})/\sqrt{2}$. The error bars are concealed as they are smaller than the data dots.}
	\label{fig:5}
\end{figure}

We can calculate the relative probabilities of two-photon counts from the measurements in the bases $(\ket{H}_{1} + \ket{V}_{1})/\sqrt{2}$ $(\ket{H}_{2} \pm e^{j\theta}\ket{V}_{2})/\sqrt{2}$ as $P_{\pm}$, to obtain the expectation value of $\langle M_{\theta}$ as $\langle M_{\theta} \rangle=P_{+}- P_{-}$. In theory, $\langle M_{\theta} \rangle$ reaches its maximum of 1 when $\theta=\Phi_O$, and in experiment we use the value $\theta$ corresponding to the  maximum value on the fixed sine curve. From the experimental results shown in Fig.~\ref{fig:5}, we obtain the extracted exchange phases $\Phi_O$ are $0.012 \pm 0.002$, $0.025 \pm 0.002$ and $0.027 \pm 0.002$ in radian for three Boson states $\ket{\mu^{\pm}}_{12}$ and $\ket{\nu^{+}}_{12}$, and $0.991\pi \pm 0.002$ in radian for the Fermion state $\ket{\nu^{-}}_{12}$, respectively. Supplemental Material~\cite{SuppMat} also shows the results for the other superposition states.

To summarize, we have for the first time systematically realized the HOM interference between the hyper-entangled two photons. It is greatly interesting to extend our work to high-dimensional region by exploring the HOM effect with hyper-entanglement in (two-dimensional) polarization and other (high-dimensional) DoF such as OAM. Benefitted from the recently developed technologies of high-dimensional entanglement preparation~\cite{Valencia2021,Valencia2020} and measurement methods~\cite{Bavaresco2018,Kong2020}, the HOM interference of such hyper-entangled two photons is expected in future. Our approach enables to directly measure the exchange symmetric and anti-symmetric phases, demonstrating an interesting and significant application of the hyper-entangled two-photon interference. More applications of the HOM interference in the more DoFs would be expected in quantum network~\cite{Wang2015, Liu2021}, quantum computation~\cite{Wang2018} and high-dimensional quantum entanglement and processing~\cite{Reimer2018, Kong2019, Erhard2020}. Our results show that expanding the Hilbert space by introducing the extra DoFs will open new opportunities for quantum information, such as turning the external and unmeasurable exchange phase in one DoF to an internal and measurable phase in another DoF. More DoFs not only contribute to high-dimensional quantum information processing~\cite{Erhard2020}, but also excite novel functions in quantum applications such as alignment-free quantum communications~\cite{D'Ambrosio2012} and complete Bell state measurement~\cite{Ecker2021}.

%%%%%%%%%%%%%%%%%%%%%%%%%%%%%%%%%%%%%%%%%%%%%%%%%%%%%%%%%%%%%%%%%%%
\begin{acknowledgments}
This work was supported by National Natural Science Foundation of China (Nos. 11922406, 12234009, 12274215); National Key R\&D Program of China (Nos. 2019YFA0308700, 2020YFA0309500); Innovation Program for Quantum Science and Technology (No. 2021ZD0301400); Program for Innovative Talents and Entrepreneurs in Jiangsu; Key R\&D Program of Guangdong Province (No. 2020B0303010001).
\end{acknowledgments}

%%%%%%%%%%%%%%%%%%%%%%%%%%%%%%%%%%%

\end{document}